# Dissociation of Quarkonium in a Strong Electric Field

A. M. Ishkhanyan[a,b] and V. P. Krainov[c,*]

[a] *Institute for Physical Research, National Academy of Sciences of Armenia, Ashratak, 0203 Armenia*
[b] *Institute of Physics and Technology, National Research Center "Tomsk Polytechnic University," Tomsk, 634050 Russia*
[c] *Moscow Institute of Physics and Technology, Dolgoprudnyi, Moscow oblast, 141700 Russia*
*\*e-mail: vpkrainov@mail.ru*

**Abstract**—The probability of tunnel decay of quarkonium (bond state of a heavy quark and a heavy antiquark) into free quarks in a strong electric field is estimated.

The interest in the properties of bound states of a pair of heavy quarks (quarkonium) has been revived in recent years. These states are generally analogous in properties to the positronium, viz., the neutral bound state of the electron and the positron. Analogous neutral bound states form two systems: charmonium $c\bar{c}$ consisting of a charmed quark and antiquark (with charges $+2e/3$ and $-2e/3$) and bottomonium $b\bar{b}$, consisting of beautiful quark and antiquark (with charges $-e/3$ and $+e/3$) [1]. The only difference is that the particles forming positronium are observable, while quarks in free form have never been observed as yet.

The exact form of the potential of the interaction between a quark and an antiquark is unknown. There exist different models that describe experimental spectra with the same accuracy. These models include, for example, the Quigg–Rosner potential [2], the Martin potential [3], and the so-called Cornell potential [4] proposed by physicists of the Cornell University (Ithaca, New York, USA). In the case of power-law potentials, the general form of these models is the trinomial

$$V = \frac{C}{r^q} + A + Br^s, \quad C < 0; \quad q, s, B > 0$$

(a logarithmic potential can be treated as the limiting case of this potential for $q, s \to 0$). In this communication, we estimate the probability of electric-field induced tunnel decay of the quarkonium into free quarks. If the electric field potential varies in space in accordance with the power law $V_E \sim r^p$ with exponent $p > s$, this field can induce the dissociation of the quarkonium with a certain probability. This occurs, for example, for the Martin potential in a constant electric field, while in the case of the Cornell potential ($q = s = 1$), the constant field cannot break the quarkonium. The corresponding experiment with the formation of free quarks in a certain region of space in which the detection is possible could serve as an argument in favor of a certain model.

Let us consider the decay of the quarkonium in the Martin model potential [3] of attraction between a heavy quark and an antiquark in their bound state (quarkonium); this potential has the well-known form

$$V(r) = A + Br^{0.1}. \quad (1)$$

Here, $A = -6.0$ GeV, $B = 6.87$ GeV, and distance $r$ between the quark and the antiquark is measured in GeV$^{-1}$ (this scale is about $10^{-3}$ fermi). These constants are determined from the experimental positions of different $s$ states of the quarkonium.

Replacing radial wavefunction $R_{n0}(r)$ of the $s$ states of quarkonium by $u_n(r)/r$, we arrive at the 1D Schrödinger equation with reduced quark mass $m_q$ and energy $E_n$ of the $s$ states ($n = 1, 2, 3, ...$):

$$-\frac{\hbar^2}{m_q}\frac{d^2 u_n}{dr^2} + V(r)u_n = E_n u_n. \quad (2)$$

In a quark–gluon plasma containing quarkonium "molecules," electric fields exist. Such fields are responsible for tunnel ionization of quarkonium leading generally (although with a rather low probability) to the formation of free quarks. This ionization is the subject of this communication.

Substituting expression (1) into (2), we obtain the Schrödinger equation in the form

$$-\frac{d^2 u_n}{dr^2} + br^s u_n = \varepsilon_n u_n,$$
$$\varepsilon_n = \frac{m_q}{\hbar^2}(E_n - A) > 0, \quad b = \frac{m_q}{\hbar^2} B \quad (3)$$

(in view of possible other applications, we introduce below an arbitrary but positive power $s$ of the potential, which is equal to 0.1 for the Martin potential).

Let us first consider the problem of determining energy $\varepsilon_n$ of quasi-classical energy levels in an unperturbed potential in zero electric field and of finding the normalization of the corresponding unperturbed quasi-classical wavefunctions. The Bohr quantization rule has the form

$$\int_0^{x_0} \sqrt{2(\varepsilon_n - br^s)}\,dr = \left(n - \frac{1}{4}\right), \quad n = 1, 2, 3, \ldots \quad (4)$$

Term 1/4 is important for low-lying states, for which the semiclassical approximation with this term is successfully applicable [5].

Change of variable $r = (z\varepsilon_n/b)^{1/s}$ reduces this integral to the beta function

$$\int_0^1 (1-z)^{1/2} z^{1/s-1} dz = B\left(\frac{3}{2}, \frac{1}{s}\right) = \frac{\sqrt{\pi}\Gamma(1/s)}{2\Gamma(3/2 + 1/s)}$$
$$= \frac{1}{\sqrt{2}} s\left(n - \frac{1}{4}\right)\pi b^{1/s} \varepsilon_n^{-1/2 - 1/s}. \quad (5)$$

Thus, semiclassical energies are given by

$$\varepsilon_n = \left(\sqrt{2\pi}\left(n - \frac{1}{4}\right)\frac{\Gamma(3/2 + 1/s)}{\Gamma(1 + 1/s)}\right)^{2s/(s+2)} b^{2/(s+2)}, \quad (6)$$
$$n = 1, 2, 3, \ldots$$

The unperturbed wavefunction in the classically accessible region has the form

$$\phi_n(r) = \frac{A}{\sqrt{p_0(r)}}\cos\left(\int_r^{r_0} p_0(r')\,dr' - \frac{\pi}{4}\right),$$
$$r_0 = \left(\frac{\varepsilon_n}{b}\right)^{1/s}. \quad (7)$$

Here, the unperturbed momentum is defined as

$$p_0(r) = \sqrt{2(\varepsilon_n - br^s)}. \quad (8)$$

Normalization factor $A$ in Eq. (7) can be determined from the condition

$$A^2 \int_0^{r_0} \frac{dr}{2p_0(r)} = 1. \quad (9)$$

Let us now evaluate the integral analogously to the previous case:

$$A^2 = \frac{2^{3/2}\Gamma(1/s + 1/2)}{\sqrt{\pi}\Gamma(1/s + 1)} b^{1/s} (\varepsilon_n)^{(s-2)/2s}. \quad (10)$$

The unperturbed quasi-classical wavefunction in the initial region under the potential barrier (behind the left point of the classical rotation), in which the electric field is still insignificant, has the form

$$\phi_n(r) \to \frac{A}{2\sqrt{|p_0(r)|}} \exp\left(-\int_{r_0}^r |p_0(r')|\,dr'\right), \quad (11)$$
$$r > r_0,$$
$$|p_0(r)| = \sqrt{2(br^s - \varepsilon_n)}. \quad (12)$$

Let us consider the main part of the subbarrier region. The quasi-classical momentum for a particle moving along the $r$ axis under the barrier has the form

$$|p(r)| = \sqrt{2(br^s - \varepsilon_n - E_0 r)}, \quad E_0 = \frac{m_q}{\hbar^2} eE. \quad (13)$$

The quasi-classical wavefunction is an analytic continuation of quasi-classical function (11) to the region in which the electric field becomes significant:

$$\phi_n(r) \to \frac{A}{2\sqrt{|p(r)|}} \exp\left(-\int_{r_0}^r |p(r')|\,dr'\right), \quad r > r_0. \quad (14)$$

The probability current density at the exit from under the barrier is precisely the ionization probability per unit time. In accordance with relation (14), it is given (with allowance for the reduced mass of the quark) by

$$w = \frac{A^2}{4}\exp\left(-2\int_{r_0}^{r_1} |p(r)|\,dr\right). \quad (15)$$

Here, $r_1$ is the point of the emergence of a particle from under the potential barrier. Since the exponent is much larger than unity, we must retain in integral (15) not only the principal term, but also the next term to obtain the correct value of the preexponential factor in the expression for the probability [6]. However, we confine our analysis to the calculation of only the principal term for obtaining estimates.

The principal term in the exponent is obtained by disregarding energy $\varepsilon_n$ of states:

$$I_0(s) = -2\int_0^{r_1} \sqrt{2(br^s - E_0 r)}\,dr,$$
$$r_1 = (b/E_0)^{1/(1-s)}. \quad (16)$$

Substitution of variable $r = (by/E_0)^{1/(1-s)}$ in this expression singles out the dependence on the field and reduces this integral the beta function also:

$$I_0(s) = -\frac{2^{3/2}}{1-s} b^{3/2(1-s)} E_0^{-(s+2)/2(1-s)}$$
$$\times \int_0^1 (1-y)^{1/2} y^{3s/2(1-s)}\,dy = -\frac{\sqrt{2\pi}}{1-s} b^{3/2(1-s)} E_0^{(s+2)/2(1-s)} \quad (17)$$
$$\times \Gamma\left(\frac{s+2}{2(1-s)}\right)\bigg/\Gamma\left(\frac{5-2s}{2(1-s)}\right).$$

We now consider the limit $s \to 0$ for the Martin potential. Expression (17) leads to

$$I_0(0) = -\frac{4\sqrt{2}}{3E_0} b^{3/2}. \qquad (18)$$

For $s \ll 1$, expressions (15) and (18) give

$$w = \sqrt{\frac{s}{2\pi}} \left(\frac{b}{\varepsilon_n}\right)^{1/s} \exp\left(-\frac{4\sqrt{2}}{3E} b^{3/2}\right). \qquad (19)$$

In this case, expression (6) for the unperturbed energy gives

$$\varepsilon_n = \left(n - \frac{1}{4}\right)^s b, \quad n = 1, 2, 3, \ldots \qquad (20)$$

Substituting this expression into (19), we obtain the following simple expression for the tunnel ionization probability per unit time:

$$w = \frac{\sqrt{s}}{\sqrt{2\pi}(n - 1/4)\tau_a} \exp\left(-\frac{4\sqrt{2}}{3E_0} b^{3/2}\right). \qquad (21)$$

Here, the characteristic "quark" time is defined analogously to the atomic time:

$$\tau_q = \frac{m_e}{m_q} \tau_a = 8.6 \times 10^{-21} \text{ s}. \qquad (22)$$

In view of the quite large exponent, we can use (returning to dimensional quantities) the elementary expression that contains not a single preexponential factor:

$$w \sim \frac{1}{\tau_q} \exp\left(-\frac{4\sqrt{2}(m_q B)^{3/2}}{3 m_q e E \hbar}\right). \qquad (23)$$

A quite different form for the tunnel exponent instead of expression (23) was proposed in [7] (formula (144)):

$$w \sim \exp\left(-\frac{2(2m\varepsilon_b)^{3/2}}{2m e E \hbar}\right). \qquad (24)$$

Here, $\varepsilon_b$ is the dissociation energy. This result corresponds to dissociation in the attractive potential that decreases with increasing distance and vanishes at infinity, but not in the attractive potential that slowly increases with the distance like the Martin potential. Indeed, in the former case, the well known (Landau–Oppenheimer) exponent contains the ionization energy of the system in an electric field, but does not contain parameters of the attractive potential. In the latter case (which was considered in our earlier publication [6]), the exponent contains only the parameter of the potential and the electric field, but does not contain the dissociation energy of the system.

Substituting the characteristic expression of electric field $eE \sim m_\pi^2 c^3/\hbar$ [7, 8] and omitting all numerical factors, we obtain from expression (23) the following relation containing only dimensionless ratios in the exponent:

$$w \sim \frac{1}{\tau_q} \exp\left[-\left(\frac{m_q}{m_\pi}\right)^{1/2} \left(\frac{B}{m_\pi c^2}\right)^{3/2}\right]. \qquad (25)$$

Substituting the masses $m_\pi c^2 = 0.135$ GeV of the pi-meson and $m_q c^2 = 1.29$ GeV of the charmed quark [9], as well as the constant $B = 6.9$ GeV of the Martin potential, we find that the exponent in expression (25) is on the order of 100. Therefore, the probability of the quarkonium decay into free quarks in a constant electric field in the Martin model is negligibly low: in accordance of relation (25), its decay occurs approximately during $10^{16}$ years, which considerably exceeds the lifetime of the Universe.

ACKNOWLEDGMENTS

The authors are grateful to V.V. Kiselev for valuable advice.

This study was supported by the Armenian State Scientific Committee (grant no. 18RF-139), Armenian National Foundation of Science and Education (grant no. PS-4986), the project "Leading Russian Research Universities (grant no. FTI_24_2016, Tomsk Polytechnic University), the Ministry of Education and Science of the Russian Federation (project no. 3.873.2017/4.6), and the Russian Foundation for Basic Research (project no. RFBR 18-52-05006).